\documentstyle{elsart}

\renewcommand{\baselinestretch}{2.0}
\begin{document}

{
\renewcommand{\baselinestretch}{1.0}
\begin{frontmatter}
\title{A Dynamical Mean Field Theory for the Study of Surface 
Diffusion Constants}
\author[Helsinki]{T. Hjelt\thanksref{Hjelt_email}}, 
\author[Helsinki,Providence]{I. Vattulainen}, 
\author[Helsinki,Providence,Jyvaskyla]{J. Merikoski},
\author[Helsinki,Providence,Tampere]{T. Ala-Nissila}, and 
\author[Providence]{S. C. Ying}

\address[Helsinki]{Helsinki Institute of Physics, 
        P.O. Box 9 (Siltavuorenpenger 20 C), FIN--00014 
        University of Helsinki, Finland}
\address[Providence]{Department of Physics, Box 1843, Brown University, 
        Providence, R.I. 02912, U.S.A.}
\address[Jyvaskyla]{Department of Physics, University of Jyv\"askyl\"a, 
        P.O. Box 35, FIN--40351 Jyv\"askyl\"a, Finland}
\address[Tampere]{Laboratory of Physics, 
        Tampere University of Technology, P.O. Box 692, 
        FIN--33101 Tampere, Finland}
\thanks[Hjelt_email]{Corresponding author. E-mail: thjelt@rock.helsinki.fi.}


\begin{abstract}
We present a combined analytical and numerical approach based on the 
Mori projection operator formalism and Monte Carlo simulations to 
study surface diffusion within the lattice-gas model.  In the 
present theory, the average jump rate and the susceptibility factor 
appearing are evaluated through Monte Carlo simulations, while the 
memory functions are approximated by the known results for a Langmuir 
gas model.  This leads to a dynamical mean field theory (DMF) for
collective diffusion, while approximate correlation effects 
beyond DMF are included for tracer diffusion. We apply our formalism to 
three very different strongly interacting systems and compare  
the results of the new approach with those of usual Monte Carlo 
simulations.  We find that the combined approach works very well for 
collective diffusion, whereas for tracer diffusion the influence of 
interactions on the memory effects is more prominent.
\end{abstract}

\end{frontmatter}
}

\pagebreak

\section{Introduction}

Most of the existing theoretical investigations of surface diffusion 
\cite{Nak80,Tah83,Cha84,Ris84,Hau87,Han90,Ala92b,Che94} 
are for cases where the interparticle interactions do not play 
an important role.  
Approximations for strongly interacting systems have only been 
developed for some special cases 
\cite{Mur79,Ree81,Maz81,Gra90,Fer93,Mys93,Tor94,Wic95,Dan95,Ala96}, 
but a general understanding of the interaction effects is still lacking. 
At finite 
coverages, there are two different diffusion constants.  The tracer 
diffusion coefficient $D_T$ is directly related to the motion of a 
tagged particle as observed in STM and field ion measurements, 
while the collective diffusion coefficient $D_C$ 
describes the macroscopic density fluctuations as measured in field 
emission and optical grating experiments \cite {Experimental}.  
Already in 1981, Reed and Ehrlich \cite{Ree81} tried to relate $D_C$ 
also to the individual adparticle motion.  They proposed that $D_C$ 
can be expressed as a product of a thermodynamical factor and an 
effective jump rate of adparticles.  However, there was no rigorous 
theoretical basis for this decomposition, in which the effect of 
dynamical correlations (memory effects) is neglected \cite{Fer94}.  
These memory effects originate from the fact that a backward jump of a 
diffusing particle is more likely than jumps to other sites.

The purpose of this Letter is to present a new method for studying 
surface diffusion within the lattice gas model \cite{Keh84}.  It is 
based on a combination of an analytic approach and Monte Carlo (MC) 
simulations.  The only approximation involved in the new approach is 
that the memory effects are approximated by the known results for the 
Langmuir gas model \cite{Gom90}.  In the Langmuir gas, there are no 
direct interactions, but the double occupancy of lattice sites is 
excluded.  It turns out that our result for $D_C$ is exactly of the form 
proposed in Ref.~\cite{Ree81}.  We apply the theory to three different 
physical systems: O/W(110) \cite{Vat96}, adlayer on a stepped 
substrate \cite{Mer96}, and flexible, chainlike molecules on a flat 
substrate \cite{Ala96}.  We find that the predictions of the theory 
for collective diffusion compare very well with full MC 
simulations, thus justifying the validity of the Reed-Ehrlich 
description \cite{Ree81}.  Moreover, the new method is computationally 
very efficient.  For tracer diffusion, however, we find the effect 
of interactions on the memory effects to be more pronounced.

\section{Theoretical description of surface diffusion}

As a starting point for studying collective diffusion, we focus on 
the density-density autocorrelation function $S(|{\bf r} - {\bf r}^{'} 
|,t) = \langle \delta n({\bf r},t) \delta n({\bf r}^{'},0) \rangle$, 
where $\delta n({\bf r},t) = n({\bf r},t) - \langle n({\bf r},t) 
\rangle$ with an occupation variable $n({\bf r},t) = 0,1$ at a lattice 
site ${\bf r}$ at time $t$.  The corresponding Laplace-Fourier 
transform is denoted by $S({\bf q},z)$.  Using the Mori projection
operator formalism \cite{Mor65}, it can be shown that \cite{Tor94}
\begin{equation}
\label{Eq:Mori}
 S({\bf q},z) = 
   \frac{ \chi({\bf q}) }
        { z - b ({\bf q})\chi({\bf q})^{-1} + M({\bf q},z) },
\end{equation}
where $b({\bf q})$ contains microscopic jump rate information, 
$\chi({\bf q})$ is the thermodynamical susceptibility, and the memory 
function $M({\bf q},z)$ contains the dynamical correlations.  The 
collective diffusion constant can be obtained from the correlation 
function $S({\bf q},z)$ by examining its pole in the limits ${\bf q 
}\rightarrow 0 $ and $z \rightarrow 0 $ \cite{Fer93,Tor94}.  In the 
limit ${\bf q }\rightarrow 0 $, $ b({\bf q}) \sim N \sum_{\alpha} 
\Gamma_{\alpha} q_{\alpha}^2 \ell_{\alpha}^2 $, where $N$ is the 
number of adparticles in a system, 
$ \Gamma_{\alpha} $ is the average jump rate and $ \ell_{\alpha} $ is 
the jump length along the direction $\alpha = x,y$.  Also $\chi_0 = 
\lim_{q\to 0} \chi({\bf q})/N = \langle (\delta N)^2 \rangle / \langle 
N \rangle $ is just the compressibility of the adsorbate overlayer.
Note that when the memory function $M$ is left out, this constitutes a 
dynamical mean field theory (DMF) for the collective diffusion constant.

Our novel combined approach consists of evaluating the average jump 
rate $ \Gamma_{\alpha} $ and the compressibility $\chi_0$ not through 
further analytic approximations \cite{Tor94}, but rather by direct MC 
simulations.  $\Gamma_{\alpha}$ is directly obtained from the success 
ratio of individual particle jumps in the canonical ensemble,
while $\chi_0 $ is more conveniently evaluated within the grand 
canonical ensemble.  This procedure is easily implemented for 
arbitrary interaction strengths and transition algorithms 
\cite{Keh84} within the lattice gas model.  For the memory function 
$M$, we follow Ferrando et al. \cite{Fer93,Tor94} and approximate it 
by the known expression for the Langmuir gas model.  Thus the effect 
of direct interparticle interactions on $M$ is not taken into 
account.  In the Langmuir gas model the memory function for collective 
diffusion is exactly zero \cite{Gom90,Kut81}, 
and our approach corresponds to the  DMF approximation for $D_{C}$.  
Eq.~(\ref{Eq:Mori}) then leads to an expression for $D_{C}$ exactly of 
the form first proposed by Reed and Ehrlich: 
\begin{equation}
D_{C,\alpha\alpha}^{{\rm appr}} = \frac{\ell_{\alpha}^2}{4} 
\frac{\Gamma_{\alpha}}{\chi_0} \ . 
\label{collective}
\end{equation}

For tracer diffusion, a similar analysis can be performed
for the self-correlation function $S_s(|{\bf r}-{\bf r}^{'} 
|,t) = \langle \delta n_s({\bf r},t) \delta n_s({\bf r}^{'},0) 
\rangle$ with $\delta n_s({\bf r},t) = n_s({\bf r},t) - \langle 
n_s({\bf r},t) \rangle$, where the tracer occupation number 
at ${\bf r}$ at time $t$, $n_s({\bf r},t) = \delta({\bf r} - {\bf R}(t))$, 
now refers to a tagged particle at ${\bf R}(t)$.
In the expression analogous to Eq.~(\ref{Eq:Mori}),  the jump 
factor $b ({\bf q})$ remains the same, $\chi({\bf q})$ is replaced by 
unity and the memory function $M_{s}({\bf q},z) $ differs from the 
corresponding one for collective diffusion. 
Again, we approximate the correlation effects resulting from  
$M_{s}({\bf q},z)$ by the known expression for the Langmuir model.  
The resulting expression for $D_{T}$ becomes \cite{Fer93} 
\begin{equation}
D_{T,\alpha\alpha}^{{\rm appr}} = 
	\frac{\ell_{\alpha}^2}{4} f(\theta) \Gamma_{\alpha} \ ,  
\label{tracer}
\end{equation}
where $f(\theta)$ is a known  correlation factor 
\cite{Nak80,Tah83,Cha84} that depends only on the coverage $\theta$ 
and the geometry of the lattice.  To summarize, our approach yields 
a DMF result for the collective diffusion constant $D_{C}$.  
For the tracer diffusion constant $D_{T}$,  
approximate correlation effects beyond the DMF are included.

\section{Results for model systems}

We now apply our formalism to study diffusion for three different 
systems based on lattice-gas models.  The first system is O/W(110) 
\cite{Vat96,Sah88}.  We study this system at a coverage of $\theta = 
0.45$ over a wide range of temperatures.  At low temperatures, the 
system is in the ordered p($2\times 1$) phase, while for $T > T_c$ it 
is disordered.  Details of the model and MC simulations of the 
diffusion coefficients are in Ref.~\cite{Vat96}.  In 
Fig.~\ref{Figure1}(a), we show the results for collective diffusion in 
an Arrhenius plot.  The agreement between the present DMF result and 
the direct MC data is remarkably good.  Furthermore, DMF works rather 
well in the strongly interacting region, i.e.\ in the ordered phase 
and even close to $T_c$, the difference being always less than 15\%.
For tracer diffusion shown in Fig.~\ref{Figure1}(b),
on the other hand, the discrepancy between the MC data and the
DMF corrected by the approximate correlation factor is more significant, 
and becomes most prominent in the ordered phase 
where the interaction effects are important.

The second system we consider is an adsorbate layer on
a substrate with equally spaced straight steps. In addition to 
repulsive nearest-neighbor interactions between the adsorbates, 
this model includes an extra binding energy at step edges, an extra barrier 
for climbing over step edges, and enhanced diffusion along step edges. 
A detailed description of the model and the formalism \cite{StepMori}
is given in Ref.~\cite{Mer96}. 
In Fig.~\ref{Figure2} we compare the theoretical results for 
collective and tracer diffusion with the direct MC simulations at a 
temperature comparable to the adsorbate interactions and the various 
substrate-induced activation barriers.  For collective diffusion, the 
DMF  gives again a very good description.  For tracer diffusion, with 
the chosen parametrization \cite{Mer96}, the component along the 
direction {\em perpendicular} to the steps is still described rather 
well by the approximate theory.
For the component along the direction {\em parallel} to step edges, 
the approximate theory fails already at relatively low coverages.  
This can be explained by 
the increased concentration at step edges due to the extra binding 
energy there, which leads to nontrivial memory effects when tracer 
particles have to make detours via terraces to pass each other.

The last system we consider,  polymer chain  molecules adsorbed on a 
flat substrate, is the most complex one. 
The chains are modelled by the two-dimensional fluctuating-bond model  
\cite{Car88}, in which the polymer segments occupy single sites on a 
lattice and  occupation of nearest and next nearest neighbor sites is 
excluded.  There are no direct interactions between the polymers in 
the model.  However, an effective entropic repulsive interaction  
exists between the chains. A more detailed description of the model 
can be found in Ref.~\cite{Ala96}. 
In Fig.~\ref{Figure3}(a), we show a comparison for the diffusion 
constants calculated from both 
the present theory and the full MC simulation as a function of 
concentration $\theta$ \cite{polymer}.  Even for this complex system, 
there is good agreement between the two approaches for the collective 
diffusion.  Again, for the tracer diffusion whose results are 
presented in Fig.~\ref{Figure3}(b), the discrepancy is significant, 
and the results from the two methods differ even qualitatively.

\section{Discussion and conclusions \label{section4}}

The three model examples above clearly illustrate that in the case of 
collective diffusion, dynamical correlations between successive jumps of 
individual particles cancel out to a large degree even for a strongly 
interacting system, and the dynamical mean field theory 
(DMF) works well.  For tracer diffusion, the memory effects depend 
more strongly on the interaction between the adparticles and the 
approximate correlation effects from the Langmuir gas model do not 
provide an adequate description.

The success of the DMF for $D_{C}$ has provided us with a better 
understanding of collective diffusion as well as a practical tool 
for its evaluation. From the theoretical viewpoint, $D_C$ in the DMF 
description is related precisely to an average jump rate of the 
individual adparticles, thus justifying 
the phenomenological assumption of Reed and Ehrlich. It provides 
a conceptual link between the individual particle jump rates, such 
as those measured in STM and field ion microscope experiments, and 
macroscopic density fluctuation measurements like field emission 
and optical grating studies. 
On the practical side, the calculation of the DMF description,
including the average jump rate and the compressibility of the
adparticles, is much easier to evaluate numerically 
compared with a direct simulation of $D_C$. 
Our preliminary results for the three model cases indicate that this
speedup is typically of the order of $25 - 100$, 
the method being most efficient for cases where large system sizes 
or unit cells are needed.

To conclude, our combined method provides a powerful 
tool for studying realistic models of complex systems.  
At present, we are employing the DMF formalism to study 
polymer chain molecules with direct interactions between them 
and surfaces with high concentration of imperfections such as 
steps, kinks and traps.

\bigskip\bigskip
{\bf Acknowledgements}

T. H. thanks the Academy of Finland, the Jenny and Antti Wihuri 
Foundation, and the Magnus Ehrnrooth Foundation for support. 
I. V. thanks the Neste Co. Foundation, the Jenny and Antti Wihuri 
Foundation, and the Finnish Academy of Sciences for support. 
J. M. is supported by the Academy of Finland and the Finnish Cultural
Foundation. This research has also been partially supported 
by a grant from the office of Naval Research (S. C. Y. and 
J. M.). Finally, computing resources of the University of 
Helsinki, Brown University, and the University of Jyv\"askyl\"a 
are gratefully acknowledged.

\begin{figure}[htb]
        \vspace{30mm}
        \caption{
	Results for (a) $D_C$ and (b) $D_T$ as a typical 
	Arrhenius plot in the O/W(110) system at $\theta = 0.45$. 
	The results of conventional MC simulations along the two 
	principal directions $x$ and $y$ are shown by open symbols, 
	while the corresponding results of the approximate theory 
	are presented by lines. The critical temperature of the 
	order-disorder phase transition is denoted by $T_c$. 
        The error bars of the MC results are roughly of the size of 
	the symbols for $D_C$ and very small for $D_T$. 
	\label{Figure1}}
\end{figure}

\begin{figure}[htb]
        \vspace{30mm}
        \caption{Diffusion constants (a) $D_C$ and (b) $D_T$ 
        as a function of concentration $\theta$ for diffusion in direction 
        perpendicular to the steps ($x$ direction) and parallel 
        to the steps ($y$ direction) in a model for a submonolayer 
	of interacting adsorbates on a stepped substrate.  
        The MC results are denoted by the open symbols,
        the approximate theory is shown by lines, and
        all data for $D_{yy}$ has been scaled by a factor of 0.02.
        The error bars of the MC results are roughly of the size of
        the symbols for $D_C$ and very small for $D_T$.
	\label{Figure2}}
\end{figure}

\begin{figure}[htb]
        \vspace{30mm}
        \caption{Results for (a) $D_C$ and (b) $D_T$ as a function of 
	the concentration $\theta$ in the model polymer system studied. 
	The MC results for diffusion coefficients are given by squares,  
	while the corresponding approximate descriptions are shown 
        by lines. 
	\label{Figure3}}
\end{figure}


\begin{thebibliography}{100}

\bibitem{Nak80}
K. Nakazato and K. Kitahara, 
Prog. Theor. Phys. {\bf 64}, 2261 (1980). 

\bibitem{Tah83}
R. A. Tahir-Kheli and R. J. Elliott, 
Phys. Rev. B {\bf 27}, 844 (1983). 

\bibitem{Cha84}
D. K. Chaturvedi, 
J. Phys. C: Solid State Phys. {\bf 17}, L449 (1984). 

\bibitem{Ris84}
H. Risken, 
{\em The Fokker-Planck Equation} 
(Springer-Verlag, Berlin, 1984). 

\bibitem{Hau87}
J. W. Haus and K. W. Kehr, 
Phys. Rep. {\bf 150}, 263 (1987). 

\bibitem{Han90}
P. H\"anggi, P. Talkner, and M. Borkovec, 
Rev. Mod. Phys. {\bf 62}, 251 (1990). 

\bibitem{Ala92b}
T. Ala-Nissila and S. C. Ying, 
Prog. Surf. Sci. {\bf 39}, 227 (1992). 

\bibitem{Che94}
L. Y. Chen and S. C. Ying, 
Phys. Rev. B {\bf 49}, 13838 (1994). 

\bibitem{Mur79}
G. E. Murch and R. J. Thorn, 
Phil. Mag. A {\bf 40}, 477 (1979). 

\bibitem{Ree81}
D. A. Reed and G. Ehrlich, 
Surf. Sci. {\bf 102}, 588 (1981). 

\bibitem{Maz81}
G. Mazenko, J. R. Banavar, and R. Gomer, 
Surf. Sci. {\bf 107}, 459 (1981). 

\bibitem{Gra90}
R. Granek and A. Nitzan, 
J. Chem. Phys. {\bf 92}, 1329 (1990). 

\bibitem{Fer93}
R. Ferrando and E. Scalas,
Surf. Sci. {\bf 281}, 178 (1993). 

\bibitem{Mys93}
A. V. Myshlyavtsev and V. P. Zhdanov, 
Surf. Sci. {\bf 291}, 145 (1993). 

\bibitem{Tor94}
M. Torri, R. Ferrando, E. Scalas, and G. P. Brivio, 
Surf. Sci. {\bf 307-309}, 565 (1994). 

\bibitem{Wic95}
T. Wichmann and K. W. Kehr, 
J. Phys.: Condens. Matter {\bf 7}, 717 (1995). 

\bibitem{Dan95}
A. Danani, R. Ferrando, E. Scalas, M. Torri, and G. P. Brivio, 
Chem. Phys. Lett. {\bf 236}, 533 (1995). 

\bibitem{Ala96}
T. Ala-Nissila, S. Herminghaus, T. Hjelt, and P. Leiderer,
Phys. Rev. Lett. {\bf 76}, 4003 (1996).

\bibitem{Experimental}
Proceedings of ``Surface Diffusion: Atomistic and
Collective Processes'' (1996).
To appear in ``NATO ASI Series B: Physics'' series.

\bibitem{Fer94}
R. Ferrando, E. Scalas, and M. Torri, 
Phys. Lett. A {\bf 186}, 415 (1994). 

\bibitem{Keh84}
K. Kehr,
in {\it Applications of the Monte Carlo Method in Statistical Physics},
edited by K. Binder (Springer, Berlin, 1984).

\bibitem{Gom90}
R. Gomer, 
Rep. Prog. Phys. {\bf 53}, 917 (1990). 
 
\bibitem{Vat96}
I. Vattulainen, J. Merikoski, T. Ala-Nissila, and S. C. Ying, 
Surf. Sci. {\bf 366}, L697 (1996). 

\bibitem{Mer96}
J. Merikoski and S. C. Ying, 
{\tt cond-mat@babbage.sissa.it} No. 9605128 (1996),
and to be published.

\bibitem{Mor65}
H. Mori, 
Prog. Theor. Phys. {\bf 34}, 399 (1965). 

\bibitem{Kut81}
R. Kutner, 
Phys. Lett. A {\bf 81}, 239 (1981). 

\bibitem{Sah88}
D. Sahu, S. C. Ying, and J. M. Kosterlitz, 
in {\em The Structure of Surfaces II}, 
edited by J. F. van der Veen and M. A. van Hove 
(Springer-Verlag, Berlin, 1988) p. 470. 

\bibitem{StepMori}
For the adlayer on a stepped substrate, 
the local concentration and jump rates vary across the terrace
so that occupation variables $n({\bf r},t)$ have several components
and $b$ and $\chi$ in Eq.~(\protect\ref{Eq:Mori}) become matrix 
quantities \protect\cite{Mer96}.
However, it is still possible to interpret the emerging terms in the 
framework of Eqs.~(\protect\ref{collective}) and (\protect\ref{tracer}).

\bibitem{Car88}
I. Carmesin and K. Kremer, 
Macromolecules {\bf 21}, 2819 (1988).

\bibitem{polymer} 
For the flexible bond model, diffusion becomes isotropic and $\Gamma$ 
can be estimated as the average success ratio of individual monomer 
jumps. The effective jump length $\ell$ has been estimated from 
the zero coverage limit. 



\end{thebibliography}
\end{document}